\documentclass[reprint,
amsmath,amssymb,
 aps,
 prb
 ]{revtex4-2}

\usepackage{graphicx}
\graphicspath{{figures/}}
\usepackage{dcolumn}
\usepackage{bm}
\usepackage[dvipsnames]{xcolor} 
\usepackage{hyperref}
\hypersetup{
           breaklinks=true,   
           colorlinks=true,   
           allcolors =BlueViolet 
        }
\usepackage{dsfont}
\usepackage{wrapfig}
\usepackage{physics}
\usepackage{subcaption}
\usepackage{cancel}
\usepackage{verbatim}

\begin{document}

\title{Long-range coupling between superconducting dots induced by periodic driving}

\author{Andriani Keliri}\email{akeliri@lpthe.jussieu.fr}
 \affiliation{Laboratoire de Physique Th\'{e}orique et Hautes Energies,
Sorbonne Universit\'{e} and CNRS UMR 7589, 4 place Jussieu, 75252 Paris Cedex 05, France}
\author{Beno\^{i}t Dou\c{c}ot}
 \affiliation{Laboratoire de Physique Th\'{e}orique et Hautes Energies,
Sorbonne Universit\'{e} and CNRS UMR 7589, 4 place Jussieu, 75252 Paris Cedex 05, France}


\begin{abstract}
We consider a Josephson bijunction consisting of three superconducting reservoirs connected through two quantum dots. In equilibrium, the interdot coupling is sizable only for distances smaller than the superconducting coherence length. Application of commensurate dc voltages results in a time-periodic Hamiltonian and induces an interdot coupling at large distances. The basic mechanism of this long-range coupling is shown to be due to local multiple Andreev reflections on each dot, followed by quasiparticle propagation at energies larger than the superconducting gap. At large interdot distances we derive an effective non-Hermitian Hamiltonian describing two resonances coupled through a continuum.
\end{abstract}

\maketitle

\section{Introduction} 
When a Josephson junction is phase biased, Cooper pairs can be transmitted through the junction, resulting in a dissipationless supercurrent \cite{Josephson,josephson_observation}. The microscopic process explaining this phenomenon is Andreev reflection, where an outgoing Cooper pair is a result of an incoming electron reflected into a hole on a normal/superconducting interface \cite{andreev1,*andreev2}. Consequently, superconducting correlations are nonzero in the normal region of the junction and Andreev bound states (ABSs) form \cite{abs,pannetier}. Moreover, when a voltage difference $V$ is applied across the junction, quasiparticles change their energy by $eV$ when traversing the normal region. The quasiparticles can then overcome the superconducting gap of energy $2\Delta$ by undergoing multiple Andreev reflections (MARs). Then, whenever the voltage is an integer subdivision of the gap, $eV=2\Delta/n$, there is an additional contribution to a dc dissipative current, resulting in a subgap structure of the current-voltage characteristics \cite{BTK,octavio,averin,subgap,cuevas,shumeiko,yeyati-dot,QDreview}. \par
For quantum dots (QDs) coupled to superconducting reservoirs (S) in the presence of voltage bias, it has been shown that the equilibrium ($V=0$) ABSs are replaced by resonances with a finite width, since MARs provide a mechanism of coupling to the continuum of states of the reservoirs \cite{fwsa,engineering,berry}. Floquet replicas of these resonances, separated by integer multiples of the drive, appear due to the time-periodicity of the system \cite{Floquet,shirley,sambe}. Therefore, superconducting quantum dots offer the unique advantage of exploring Floquet physics without suffering from thermalization problems \cite{lazarides}. Indeed, some mechanism of energy localization is required to avoid thermalization \cite{PhysRevLett.116.250401}. Here, this is provided by the superconducting gap and the fact that the ABSs of quantum dots remain detached from the superconducting continua. This in turn produces sharp Floquet resonances when the voltage is turned on, provided the coupling to the superconductors is small with relation to the superconducting gap.\par
In multiterminal configurations, commensurate voltages are required for having a single basic frequency in the system. The simplest nontrivial case then involves a three-terminal junction biased in the quartet configuration of voltages, where two superconductors are biased at opposite voltages $V_a=-V_b,$ and the third one is grounded, $V_c=0.$ Besides the general interest in multiterminal Josephson junctions as synthetic topological matter \cite{riwar2016multi}, the quartet configuration is of interest since it permits a dc supercurrent and correlations between Cooper pairs \cite{quartets1,quartets2,quartets3,quartets4}. \par
In the case of a three-terminal S-QD-S-QD-S junction, which we will also call a bijunction, and in the absence of voltage bias, the ABSs on each dot hybridize and form an Andreev molecule, producing nonlocal effects in the Josephson current. The Andreev molecule and its signatures have been the recent subject both of theoretical \cite{molecule,scattering,barakov1,barakov2,moleculedots} as well as of experimental studies \cite{InAs1,InAs2,moleculeexperimental}. When the Andreev molecule is biased in the quartet configuration, the molecular character of the system causes splitting of the Floquet resonances and modification of the subgap structure \cite{previouspaper}. Moreover, in contrast to the equilibrium case, one expects that a nonlocal coupling between the dots of the biased system should persist at distances much larger than the superconducting coherence length $\xi_0$ \cite{melin2021}. We have previously shown that at large interdot distances, the system behaves like an interferometer, resulting in a subgap current that oscillates as a function of the voltage \cite{previouspaper}. The interference is due to a Floquet version of the geometrical interference effect first discovered by Tomasch in thick superconducting films \cite{Tomasch1,*Tomasch2,Tomasch3}. The Tomasch effect ensues from the interference between electronlike and holelike quasiparticles which are degenerate in energy, but differ in their wavenumbers $k_{e,h}-k_F=\pm\sqrt{E^2-\Delta^2}/\hbar v_F$ \cite{McMillan-Anderson,Wolfram}. As a result of the interference, the tunneling current and the density of states (at energies larger than the gap) oscillate as a periodic function of the applied voltage $V$ and the thickness $d$ of the film that appear in the combination $\frac{2d}{\hbar v_F}\sqrt{(eV)^2-\Delta^2}$. A typical thickness in the Tomasch experiments was a few tens of micrometers, which corresponds to a distance two orders of magnitude larger than a typical superconducting coherence length. \par
In this paper, we show that the long-range coupling between the dots of the driven bijunction is due to processes that involve local MARs on each dot, followed by quasiparticle propagation at energies above the gap $\abs{E}>\Delta$ in the middle superconductor, in agreement with \cite{melin2021}. We focus on the consequences of this Floquet-Tomasch effect on the spectrum of the bijunction, particularly in the subgap region $\abs{E}<\Delta,$ and find that oscillations appear, superimposed on the single junction spectrum. The corresponding pole structure of the resolvent is drastically modified with respect to the resolvent of the single junction, and the number of poles found increases with the dot separation. We show that the modification of the resolvent around the single junction resonances, as well as the resulting oscillations in the spectrum, can be accounted for by deriving an effective non-Hermitian two-level model of resonances coupled through a continuum. The continuum in this case acts as the sole source both of dissipation and of coupling.\par
The rest of the paper is organized as follows: in Sec. \ref{model} we present the model Hamiltonian and map the problem to a tight-binding chain with sites labeled by Floquet modes. We discuss the coupling of the two dots at the limit of large interdot distances. In Sec. \ref{TLS} we derive an effective Floquet Hamiltonian corresponding to two discrete states coupled through a superconducting continuum. Conclusions are presented in Sec. \ref{conclusions}. Details on the derivation of the effective two-level Floquet operator are presented in Appendix \ref{diagonalization}.

\section{Model and method}\label{model}
\begin{figure}[t]
    \includegraphics[width=0.6\linewidth]{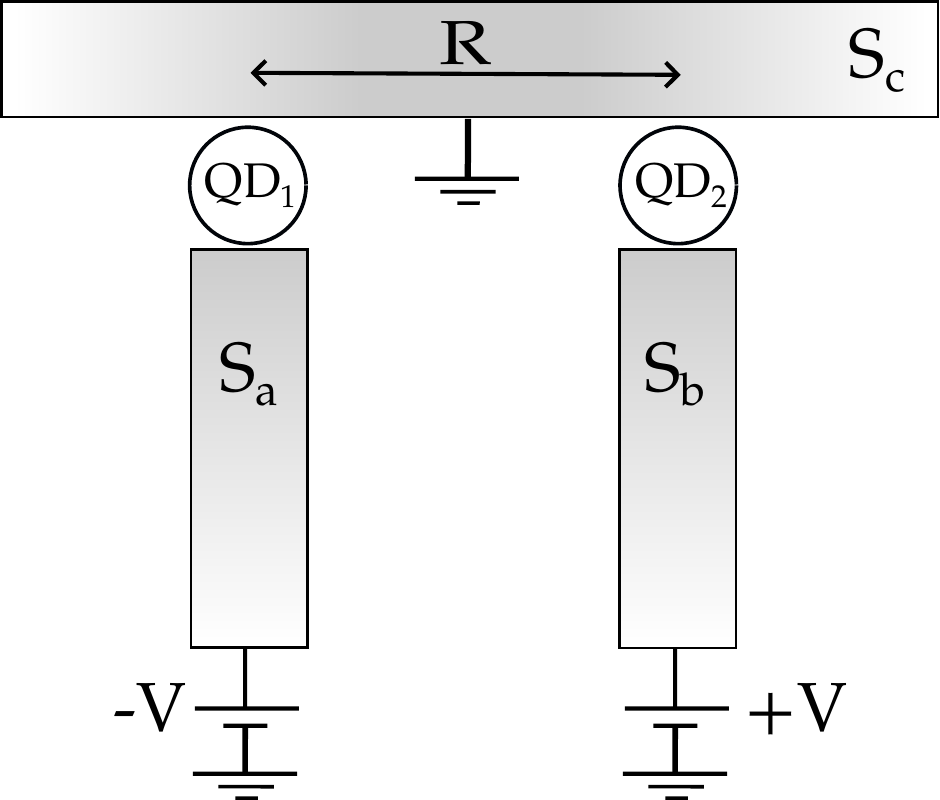}    
    \caption{Sketch of the three-terminal junction considered in this paper.}
    \label{sketch}
\end{figure}
\subsection{Hamiltonian}
We consider a Josephson bijunction as depicted in Fig. \ref{sketch}, composed of three superconducting reservoirs and two quantum dots. For simplicity, the quantum dots are modeled by discrete levels at zero energy. The reservoirs being biased with commensurate dc voltages, the resulting Hamiltonian is time-periodic, and Floquet theory can be applied. The configuration $(V_a,V_c,V_b)=(-V,0,+V)$ used here means that a basic frequency $\omega_0=eV/\hbar$ exists in the system. Moreover, using the Josephson relation $\dot{\phi_j}(t)=2eV_j/\hbar,$ we see that this choice of voltages leads to a static phase $\phi_q=\phi_a(t)+\phi_b(t)-2\phi_c,$ where $\phi_q$ is called the quartet phase \cite{quartets1}. Without loss of generality, we can choose a gauge where $\phi_c=0.$ \par
The total Hamiltonian of the bijunction is
\begin{equation}
    \mathcal{H}(t)= \mathcal{H}_0 +\mathcal{V}(t),
\end{equation} 
where the static part $\mathcal{H}_0$  is a sum of BCS Hamiltonians describing the superconducting reservoirs $j=\{a,b,c\}$: 
\begin{equation}\label{bcs}
\begin{split}
    \mathcal{H}_0 &= \sum_{jk\sigma}\epsilon_{k} c^{\dagger}_{jk\sigma} c_{jk\sigma} \\
    &+\sum_{jk} \qty(\Delta e^{i\phi_j} c^{\dagger}_{jk\uparrow} c^{\dagger}_{j-k\downarrow} +\Delta e^{-i\phi_j} c_{j-k\downarrow} c_{jk\uparrow}),
\end{split}
\end{equation}
and the time-dependent part $\mathcal{V}(t)$ describes the tunneling between dots labeled by $i=\{1,2\}$ and reservoirs labeled by $j:$
\begin{equation}\label{tunneling}
 \mathcal{V}(t) =\sum_{i\in\mathrm{dots}}\sum_{jk\sigma} ( J_j(x_i) e^{is_j\omega_0 t} d^{\dagger}_{i\sigma}c_{jk\sigma} +\mathrm{h.c.}).
\end{equation}
The operators $c^{(\dagger)}_{jk\sigma}$ create (annihilate) an electron in the $j$ reservoir with momentum $k$ and spin $\sigma,$ while corresponding operators on the dots are denoted by $d^{(\dagger)}.$ For convenience, we take the dots' positions to be at $x_1=0,x_2=R,$ and the tunnel couplings to be $J_j(x_i)=J_je^{ikx_i},$ with a real amplitude $J_j=J_j^{\ast}.$ We have moreover used the notation $V_j=s_jV.$

\subsection{Mapping to a tight-binding chain}
Using the basic idea of the Floquet method \cite{shirley,sambe,hanggi,floquetengineering}, quantities can be expanded into Fourier modes $e^{-im\omega_0t},$ where integers $m$ can be thought of as positions on a fictional Floquet direction. One then obtains a time-independent tight-binding model in an extended Hilbert space \cite{shirley,grempel}. A common procedure is to `project out' the contribution of sites $n\neq m$ up to some large Floquet index $n=N,$ and arrive to an effective Floquet Hamiltonian for the site $m$ \cite{floquetengineering}. The dimensions of the obtained tight-binding model depend on the number of incommensurate drive frequencies \cite{PhysRevX.7.041008}. Here, we have one basic frequency across the system, so we will obtain an effective 1D tight-binding model. \par
The main idea is that since the system does not thermalize, we can still use the notion of quasiparticle. We therefore start by constructing dressed quasiparticle operators $\Gamma^{\dagger}(t)$ \cite{previouspaper,berry}  which are time-periodic solutions of the Bogoliubov--de Gennes (BdG) equations:
\begin{equation}\label{BdG}
    i\dv{t}\Gamma^{\dagger}_{\sigma}(t)=\comm{\mathcal{H}(t)}{\Gamma^{\dagger}_{\sigma}(t)},
\end{equation}
and therefore obey the Floquet theorem 
\begin{equation}
    \Gamma^{\dagger}(t+T)=e^{-iEt}\Gamma^{\dagger}(t).
\end{equation} 
\begin{widetext}
Here, $E$ is the quasienergy, defined modulo the frequency of the drive \cite{zeldovich}. Written as a Fourier series, the creation operator is:
\begin{equation}\label{qp operator}
 \Gamma^{\dagger}_{\sigma}(t) = \sum_{m\in\mathbb{Z}} e^{-i(E+m\omega_0)t} \Bigg[ \sum_{i\in\mathrm{dots}}\qty(u_m(i)d^{\dagger}_{i\sigma}+\sigma v_m(i)d_{i-\sigma}) 
 + \sum_{jk}\qty(u_m(jk)c^{\dagger}_{jk\sigma}+\sigma v_m(jk)c_{j-k-\sigma})\Bigg],     
\end{equation}
where $u_m(i),v_m(i)$ are the electronlike and holelike amplitudes on the dot $i.$ By plugging Eq.(\ref{qp operator}) into Eq.(\ref{BdG}) and integrating out the amplitudes of the reservoirs $u_m(jk),v_m(jk),$ we arrive at a set of eigenvalue equations for the amplitudes on the dots:
\begin{equation}\label{FLS}
	\begin{split}
		(E+m\omega_0) u_m(i) & = \sum_{ji'}\bqty{g_{j,ii'}^{11}\pqty{m+s_j}u_m(i')+g_{j,ii'}^{12}\pqty{m+s_j}v_{m+2s_j}(i')}\\
		(E+m\omega_0)v_m(i) & = \sum_{ji'}\bqty{g_{j,ii'}^{21}\pqty{m-s_j}u_{m-2s_j}(i')+g_{j,ii'}^{22}\pqty{m-s_j}v_m(i')}.
	\end{split}
\end{equation}
\vspace{-10pt}
\end{widetext}
The above equations involve `local' Green's functions $g_{j,ii'}\delta_{ii'}\equiv g_j$ for the 1D superconductor of the $j$ reservoir, defined here as
\begin{equation}\label{greens functions}
\begin{split}
	g_j(\omega) &= \frac{\Gamma_j}{iv_F q(\omega)} \mqty(\omega & -\Delta e^{i\phi_j} \\ -\Delta e^{-i\phi_j} & \omega), \qand \\
	v_F q(\omega) &\equiv i\sqrt{\Delta^2-\omega^2} \theta(\Delta-\abs{\omega}) \\ &+ \mathrm{sign}(\omega)\sqrt{\omega^2-\Delta^2} \theta(\abs{\omega}-\Delta),
\end{split}
\end{equation}
as well as a nonlocal Green's function $g_{j,ii'}(\omega)(1-\delta_{ii'})\equiv g_j(\omega,R)$ which couples the two dots:
\begin{equation}\label{nonlocalgreens}
g_j(\omega,R) = e^{iq(\omega) R}\bqty{\cos(k_F R) g_j(\omega) +\sin(k_F R) \Gamma_j \sigma_z},
\end{equation}
where $k_F$ is the Fermi wavevector. The phase $k_F R$ will be assumed fixed in order to avoid rapid oscillations at the Fermi wavelength scale . We have used the notation $\Gamma_j=\pi\rho_0 J_j^2,$ where $\rho_0$ is the density of states in the normal state of the superconductors. Moreover, we are using the shorthand $\bullet(m)\equiv\bullet(E+m\omega_0)$ in order to lighten the notation.\par 
The only nonlocal Green's function is for $j=c$ since $S_c$ is the only reservoir that couples with both dots. Due to the factor $e^{iq(\omega)R},$ the Green's function $g_c(\omega,R)$ decays exponentially at distances larger than $\xi_0$ for energies inside the gap $\abs{\omega}<\Delta,$ while for energies outside the gap $\abs{\omega}>\Delta,$ it oscillates without decay as long as there is no mechanism of decoherence in $S_c$. A finite quasiparticle lifetime \cite{dynes} will eventually produce decay of the quasiparticle propagation in $S_c$ over a mesoscopic coherence length that should be between two to three orders of magnitude larger than $\xi_0$ \cite{melin2021}. \par
We rewrite Eq. (\ref{FLS}) on the basis of the Nambu spinor $\Psi_m\equiv(u_m(1), v_m(1), u_m(2), v_m(2))^T$ which collects the amplitudes on the two dots, by defining a linear operator $\mathcal{L}$ that acts on the states $\Psi_m:$
\begin{equation}\label{tb}
(\mathcal{L}\Psi)_m \equiv M^0_m\Psi_m -M^+_{m+1}\Psi_{m+2}-M^-_{m-1}\Psi_{m-2}=0.
\end{equation}
Equation (\ref{tb}) defines a `Floquet chain operator' $\mathcal{L}$. Written in a matrix representation, it is a tridiagonal block-matrix of dimension $\mathrm{dot}\otimes\mathrm{Nambu}\otimes\mathrm{Floquet}.$ In the tight-binding analogy, the matrix $M^0_m$ describes an on-site energy at position $m$ of the chain, while matrices $M^{\pm}_{m\pm 1}$ describe hopping to neighboring sites through local Andreev reflections. The recursive character of Eq. (\ref{tb}) makes it possible to write the Floquet chain operator in a continued fraction form \cite{tridiagonal,hanggi}: 
\begin{widetext}
\vspace{-5pt}
\begin{subequations}\label{chain}
\begin{align}
    \mathcal{L}_{mm} &\equiv\mathcal{L}(m) =M^0(m)-\Sigma^{+}(m)-\Sigma^{-}(m),\\
    \Sigma^{+}(m) &= M^{+}(m+1)\frac{1}{M^0(m+2)-\Sigma^{+}(m+2)}M^{-}(m+1),\\
    \Sigma^{-}(m) &=M^{-}(m-1)\frac{1}{M^0(m-2)-\Sigma^{-}(m-2)}M^{+}(m-1).
\end{align}
\end{subequations}
\vspace{-5pt}
\end{widetext}
The explicit form of the matrices in Eq. (\ref{chain}) will be discussed in the following sections. Throughout this paper, we will concentrate on the diagonal part $\mathcal{L}_{00}=\mathcal{L}(0),$ since the zeroes of $\det\mathcal{L}_{mm}$ correspond to the eigenvalues of an effective Floquet Hamiltonian for the site $m$, and therefore give access to a Floquet spectrum. In fact, if we introduce the resolvent operator $\mathcal{R}$ defined as the inverse of the operator $\mathcal{L},$ then the spectral function can be found by taking an appropriate trace of the resolvent operator in the Nambu subspace of one of the dots \cite{thesispillet}. More precisely, a time-averaged spectral function over one period of the drive can be defined \cite{positivity} as proportional to the imaginary part of the resolvent operator in the subspace of one of the dots. If, for example, a normal probe is tunnel coupled to dot $1,$ the spectral function will be given by:
\begin{equation}\label{spectral function}
    \mathcal{A}_1(\omega) =-\frac{1}{\pi}\Im\bqty{\mathcal{R}^{11}(\omega)+ \mathcal{R}^{22}(-\omega)}=-\frac{2}{\pi}\Im\mathcal{R}^{11}(\omega).
\end{equation} 
Expressions for the non-diagonal parts of $\mathcal{L}$, needed for calculating more complicated observables like the current, were given in previous work \cite{previouspaper}. Equation (\ref{chain}) can be seen as a Dyson equation, with self-energy matrices $\Sigma^{\pm}$ that renormalize the zeroes of $M^0$ by adding a finite imaginary part to them. This imaginary part is introduced in practice by the Green's functions, contained in the self-energy, which become imaginary at energies larger than the gap $\abs{\omega}>\Delta.$ Physically, this corresponds to coupling the initial discrete levels (the ABSs) on the dot(s) to the superconducting continua through MAR. Then, $\Sigma^{+}$ corresponds to MAR processes which raise the energy of a quasiparticle above the gap $\omega>\Delta,$ while $\Sigma^{-}$ corresponds to MAR processes which lower the energy below the gap $\omega<-\Delta.$ Technically, one can truncate the continued fractions at some cutoff index $\abs{N}> \frac{\Delta}{\omega_0}$ by considering that the self-energies become small $\Sigma^{\pm}(\pm N)\to 0$ at large energies $\abs{\omega\pm N\omega_0}\gg \Delta.$ Therefore, at voltages which are a significant fraction of the gap, one can greatly simplify the expressions of $\Sigma^{\pm},$ while at small voltages an increasingly greater number of Floquet harmonics need to be taken into account. We will here concentrate on the former regime, since it facilitates the analytical part of the work while giving some insight on the involved mechanism of coupling. However, the Floquet-Tomasch mechanism of coupling that will be described in the next section occurs at smaller voltage values as well, albeit at higher MAR order and therefore at a higher order in the tunnel couplings.

\begin{figure}
    \centering
    \begin{subfigure}{0.8\linewidth}
    \includegraphics[width=\linewidth]{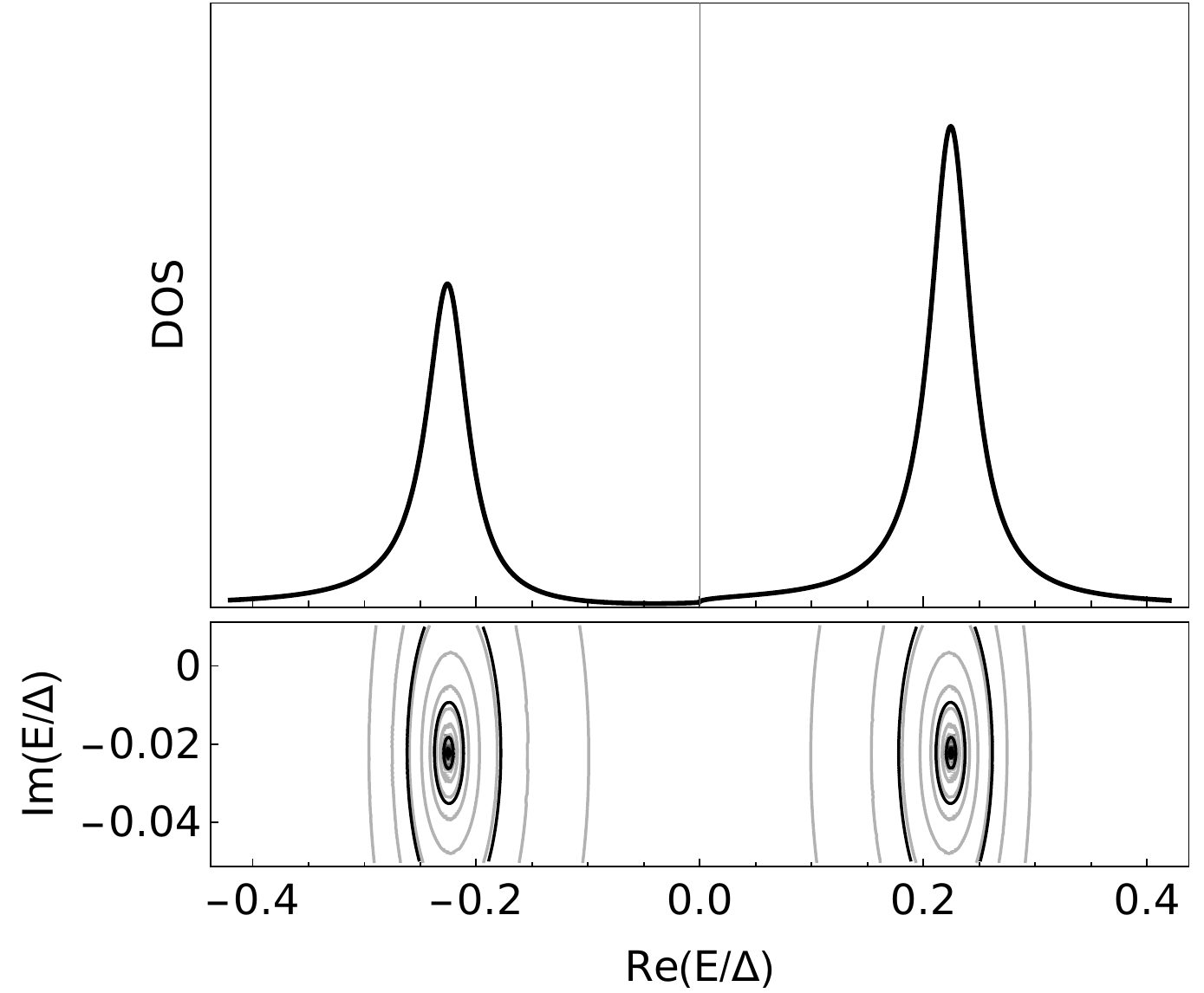} 
    \caption{}
    \label{dos_single}
    \end{subfigure}
   \begin{subfigure}{0.8\linewidth}
    \includegraphics[width=\linewidth]{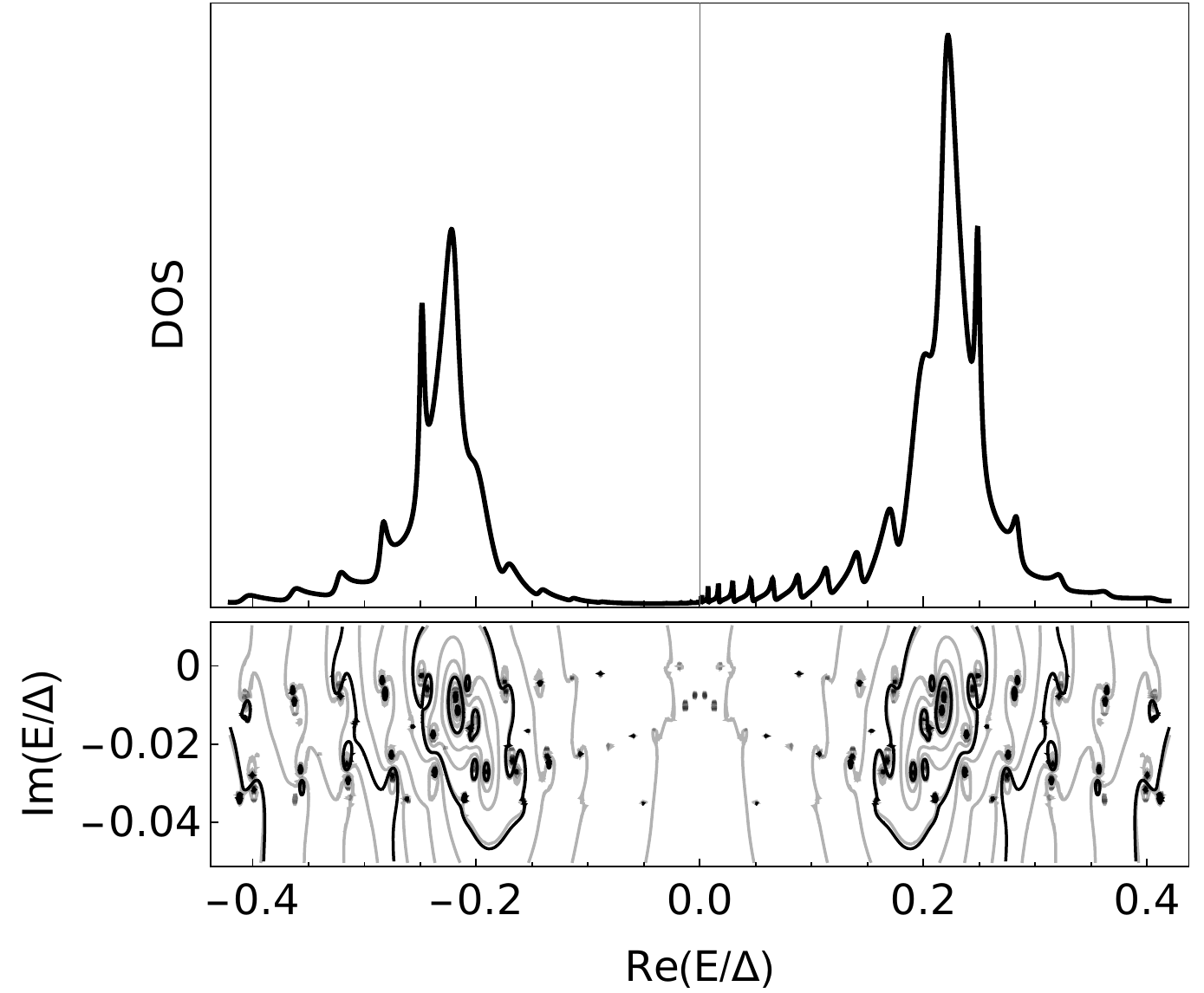}
    \caption{}
    \label{dos_bijunction}
   \end{subfigure}
    \caption{Contour plot of $\log(\abs{\det\mathcal{L}(0)})$ in the complex plane showing the zeroes of $\det\mathcal{L}(0)$ (lower panel) and the corresponding spectral function of dot $1,$ $-\frac{2}{\pi}\Im\mathcal{R}^{11}$ (upper panel). (a) Single junction. (b) Bijunction when the distance between the dots is $R=50\xi_0,$ showing the interference effect in the DOS due to Floquet-Tomasch processes. All couplings are set to $\Gamma_j=\Delta/2$ and the frequency of the drive is $\omega_0=\Delta/2.$}
    \label{dos}
\end{figure}
\subsection{Large voltage bias, large separation approximation}
\begin{figure*}
    \centering
    \begin{subfigure}{\linewidth}
      \includegraphics[width=\linewidth]{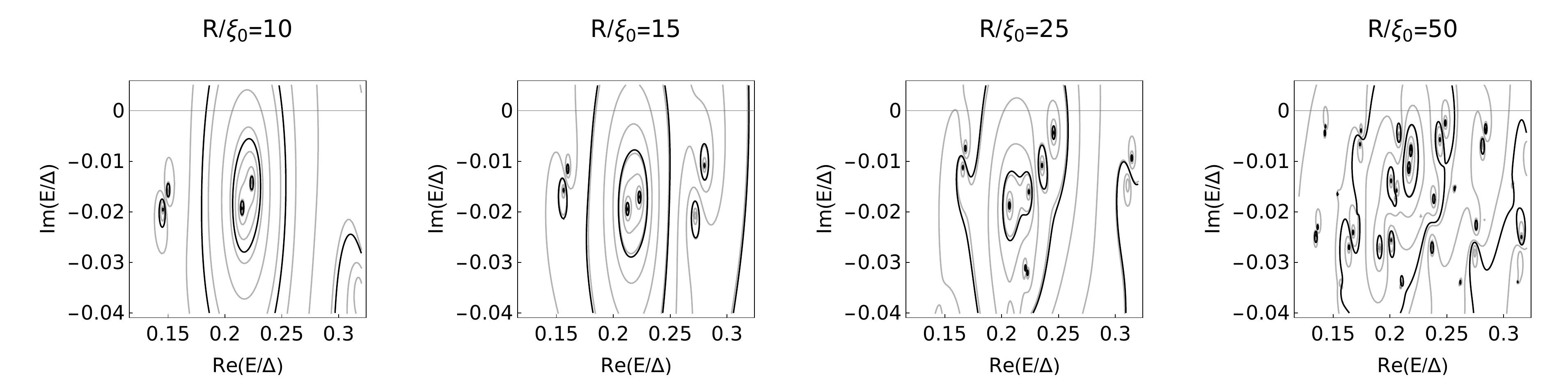}  
      \caption{}
    \end{subfigure}
    \begin{subfigure}{\linewidth}
    \includegraphics[width=\linewidth]{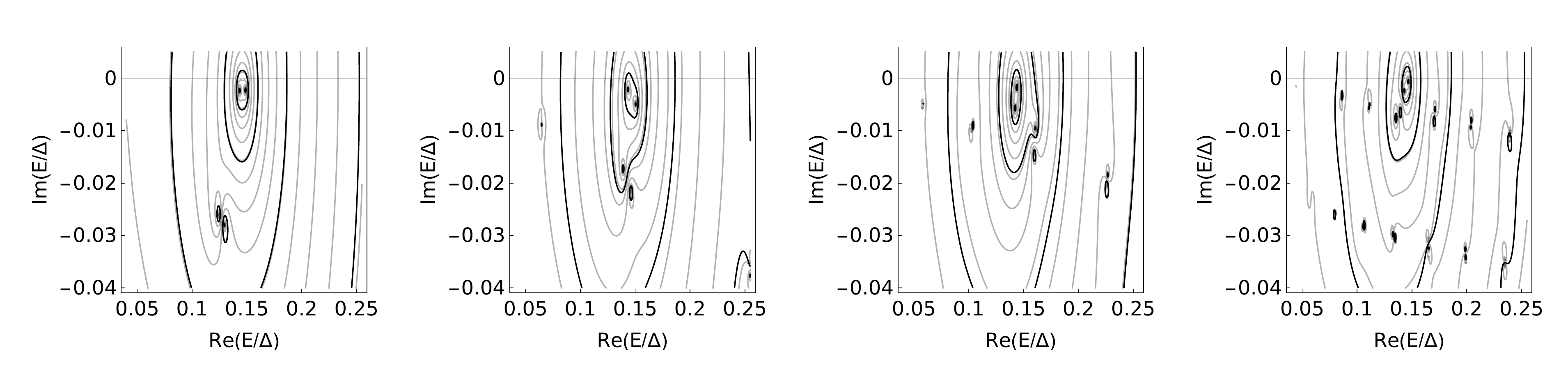}
    \caption{}
    \end{subfigure}
    \caption{Evolution of the zeroes of $\det\mathcal{L}(0)$ for different couplings to the reservoirs when the interdot distance is increased. The frequency of the drive is set to $\omega_0=\Delta/2$ and the couplings are (a) $\Gamma_j=\Delta/2,$ and (b) $\Gamma_j=\Delta/5.$}
    \label{evolution}
\end{figure*}   
We will study the bijunction in the regime of large separation between the two dots ($R\gg\xi_0$) and voltages which are a significant fraction of the gap $\frac{\Delta}{2}<\omega_0<\Delta.$ In particular, we will study the spectrum around energies close to the middle of the superconducting gap. The opposite regime of small separation ($R\lesssim\xi_0$) which results in strong hybridization of the states on the dots (molecular regime) has been studied in previous work \cite{previouspaper}. In the same work, we have moreover shown that, for energies above the superconducting gap $\abs{E}>\Delta$ and in the large separation regime $R\gg\xi_0$, the density of states (DOS) exhibits oscillations as a function of the energy and the distance due to the Floquet-Tomasch effect. \par
At equilibrium, there are two competing mechanisms for the coupling of the two dots in the molecular regime: a) crossed Andreev reflection (CAR) processes, involving the Andreev reflection of two electrons, one from each dot, which then form a Cooper pair in the middle superconductor, and b) elastic cotunneling (EC) processes, involving normal transmission of quasiparticles through the middle superconductor \cite{CAR,nonlocalAR}. In terms of the superconducting Green's functions, CAR corresponds to the anomalous propagators, while EC corresponds to the normal components. An efficient way to tune the rate between these two processes has recently been proposed and demonstrated \cite{liu2022tunable,dvir2023realization}. At equilibrium, separating the two dots at distances larger than $\xi_0$ will result in trivially recovering the spectrum of two single dots, as both CAR and EC will be exponentially suppressed. However, we will show that when the system is periodically driven, a long-range coupling develops between the dots.\par
The numerical results for the spectral function on dot $1$ of the bijunction are presented in Fig. \ref{dos_bijunction} and are compared with the spectrum of a single junction (dot $2$ decoupled from dot $1$) in Fig. \ref{dos_single}. In the single-junction case, the zeroes of the Floquet chain operator $\det\mathcal{L}(0)$ are slightly shifted below the real axis (lower panel), giving a corresponding finite width to the peaks of the spectral function (upper panel). More details on the single-junction case are presented in the Supplemental Material. In the bijunction case, the real part of the resonances is not shifted with respect to the single-junction peaks, but oscillations appear, superimposed on the single-junction peaks due to coupling with the second dot. In the complex plane, the resulting behavior is a proliferation of the zeroes of $\det\mathcal{L}(0).$ The frequency of oscillations of the resolvent and, correspondingly, the number of zeroes in the complex plane increase with the distance. The behavior of the zeroes of $\det\mathcal{L}(0)$ is shown in Fig. \ref{evolution}. At this stage, both Fig. \ref{dos} and \ref{evolution} are calculated without making any approximations, i.e. by using Eq. (\ref{chain}) and truncating the continued fractions at a large cutoff index. \par
The starting point for understanding the results of Fig. \ref{dos} and \ref{evolution} is Eq. (\ref{chain}), which at $m=0$ gives:
\begin{equation}
\begin{split}
    \mathcal{L}(0) &=M^0(0)-\Sigma^{+}(0)-\Sigma^{-}(0) \\
    &=\mqty*(M^0_1(0) & g_c(0,R) \\ g_c(0,R) & M^0_2(0))-\Sigma^{+}(0)-\Sigma^{-}(0).
\end{split}
\end{equation}
The operator $\mathcal{L}$ is written on the basis of the four-component Nambu spinor $\Psi_m$ and is therefore a $4\times 4$ matrix acting in $\mathrm{dot}\otimes\mathrm{Nambu}$ space. The diagonal blocks of $\mathcal{L}$ correspond to intradot processes, while the off-diagonal blocks correspond to interdot processes. Specifically, the dots $1$ and $2$ are each coupled by local reflections to their closest reservoirs. This information is contained in the block matrices:
\begin{equation}
\begin{split}
     M^0_{1,2}(m)= &(E+m\omega_0) \mathds{1}_2-g_c(m) \\ &-\mqty(g_{a,b}^{11}(m+s_{a,b}) & 0 \\ 0 & g_{a,b}^{22}(m-s_{a,b})).
\end{split}
\end{equation}
\begin{figure*}
 \centering
    \begin{subfigure}{0.4\linewidth}
    \includegraphics[width=\linewidth]{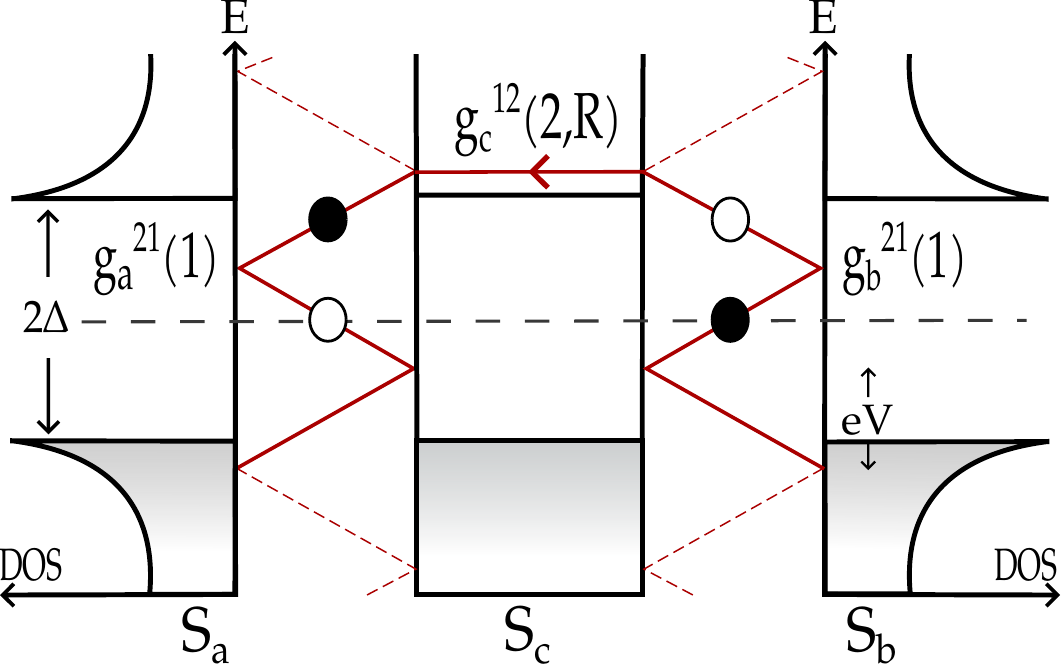} 
    \caption{}
    \end{subfigure}
     \hspace{1.2cm}
   \begin{subfigure}{0.4\linewidth}
    \includegraphics[width=\linewidth]{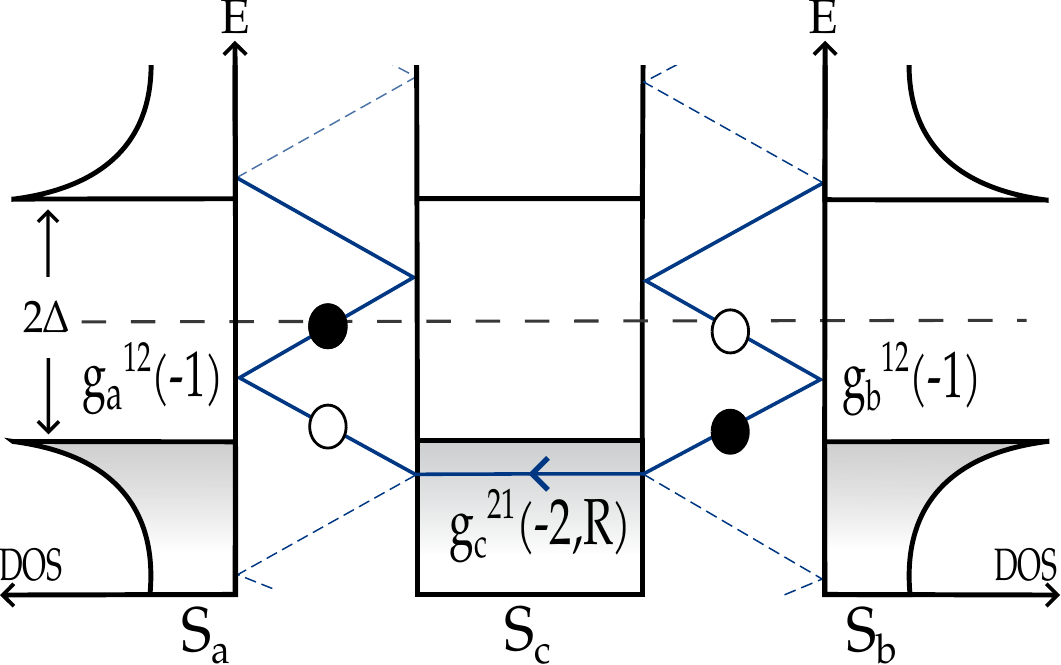}
    \caption{}
   \end{subfigure} 
   \caption{Self-energy terms (a) $\Sigma^{+,23}$ and (b) $\Sigma^{-,14}$ couple a hole (electron) on dot 1 to an electron (hole) on dot 2 through propagation in the middle reservoir. An overall phase $e^{\mp i\phi_q}e^{iq(\pm 2)R}$ is accumulated.}
    \label{selfenergies}
\end{figure*}
The off-diagonal blocks of $\mathcal{L}(0)$ couple the two dots through processes involving nonlocal Andreev reflections. The off-diagonal coupling term in $M^0(0)$ is the nonlocal Green's function of the middle reservoir $g_c(0,R)$ which is the dominant source of coupling at small distances $R\lesssim\xi_0,$ but becomes exponentially small at large distances for processes inside the gap (i.e. for energies $\abs{E}<\Delta$). Therefore, in the regime of interest $R\gg \xi_0$, the coupling of the two dots will be contained \emph{entirely} in the self-energy matrices $\Sigma^{\pm}(0).$ The self-energy elements do not go to zero as $e^{-R/\xi_0},$ but are limited by a mesoscopic coherence length instead \cite{melin2021}. For a large voltage bias $\frac{\Delta}{2}<\omega_0<\Delta,$ we can truncate the expressions for the self-energies (\ref{chain}b-c) such that $\Sigma^{\pm}(\abs{m}\geq 2)\to 0.$ Then, the self-energy matrices have the form:
\begin{widetext}
    \begin{subequations}\label{self-energy}
\begin{align}
\Sigma^{+}(0)&=\mqty*(0 & 0 & 0 & 0 \\0 & g^{21}_a(1)\bqty{\frac{1}{M^0(2)}}^{11}g^{12}_a(1)& g^{21}_a(1)\bqty{\frac{1}{M^0(2)}}^{14}g^{21}_b(1)&0 \\ 0&g^{12}_b(1)\bqty{\frac{1}{M^0(2)}}^{41}g^{12}_a(1)&g^{12}_b(1)\bqty{\frac{1}{M^0(2)}}^{44}g^{21}_b(1)&0 \\ 0 & 0 & 0 & 0), \\
\Sigma^{-}(0)&=\mqty*(g^{12}_a(-1)\bqty{\frac{1}{M^0(-2)}}^{22}g^{21}_a(-1) & 0 & 0 & g^{12}_a(-1)\bqty{\frac{1}{M^0(-2)}}^{23}g^{12}_b(-1) \\0 & 0& 0 &0 \\ 0&0&0&0 \\ g^{21}_a(-1)\bqty{\frac{1}{M^0(-2)}}^{32}g^{21}_b(-1) & 0 & 0 & g^{21}_b(-1)\bqty{\frac{1}{M^0(-2)}}^{33}g^{12}_b(-1)).
\end{align}
\end{subequations}
\end{widetext}
By inverting the $4\times 4$ matrix $M^0(\pm 2),$ one can express the self-energies (and therefore the coupling between the dots) as a function of local and nonlocal Green's functions of the reservoirs. The inversion of the matrix $M^0(m)$ can be performed blockwise. If the matrix has the form:
\begin{equation}
    M^0(m)=\mqty*(M^0_1(m) & g_c(m,R) \\ g_c(m,R) & M^0_2(m)),
\end{equation}
then we can decompose its inverse as (suppressing the indices $m,R$ for brevity)
\begin{equation}
      \frac{1}{M^0} \approx \mqty*(\frac{1}{M^0_1}+\frac{1}{M^0_1}g_c\frac{1}{M^0_2}g_c\frac{1}{M^0_1} & -\frac{1}{M^0_1}g_c\frac{1}{M^0_2} \\ -\frac{1}{M^0_2}g_c\frac{1}{M^0_1}& \frac{1}{M^0_2}+\frac{1}{M^0_2}g_c\frac{1}{M^0_1}g_c\frac{1}{M^0_2}),
\end{equation}
where we have made a perturbative expansion in the tunnel couplings and kept only terms up to $\order{\Gamma_c^2}.$ The non-diagonal terms of the self-energy can then be written as
\begin{subequations}\label{nondiag}
\begin{align}
\Sigma^{-,14} \approx & -g^{12}_a(-1)g^{21}_c(-2,R)g^{12}_b(-1) \nonumber\\
            &\times\bqty{\frac{1}{M^0_1(-2)}}^{22}\bqty{\frac{1}{M^0_2(-2)}}^{11},\\
 \Sigma^{+,23} \approx & -g^{21}_a(1)g^{12}_c(2,R)g^{21}_b(1)\nonumber\\
 &\times\bqty{\frac{1}{M^0_1(2)}}^{11}\bqty{\frac{1}{M^0_2(2)}}^{22}.
\end{align}
\end{subequations} 
The inverse processes $\Sigma^{-,41},\Sigma^{+,32}$ can be similarly obtained. The above formulas can be interpreted as specific physical processes that couple the two dots through local and nonlocal Andreev reflections. Both processes couple an electron (hole) at initial energy $\abs{E}\ll \Delta$ on dot 1 to another hole (electron) at energy $E$ on dot 2. Initially the quasiparticle on dot 2 is Andreev reflected locally on reservoir $S_b$ whereby its energy is changed by $E\pm\omega_0.$ This is then followed by a nonlocal Andreev reflection through the middle superconductor $S_c$ at energies which are above the gap $\abs{E\pm 2\omega_0}>\Delta,$ so that the propagation is \emph{not} limited by the superconducting coherence length. Finally, a local Andreev reflection on reservoir $S_a$ returns the quasiparticle to the initial energy $E$ on dot 1. A graphical representation of Eq. (\ref{nondiag}) is sketched in Fig. \ref{selfenergies}. The coupling due to processes like the above involves three Andreev reflections, meaning it is of order $\order{\Gamma_a\Gamma_c\Gamma_b}$ in the tunnel couplings. We can also see that the three Andreev reflections will contribute a quartet phase factor $e^{\pm i\phi_q},$ where $\phi_q=\phi_a+\phi_b-2\phi_c, \phi_c=0.$ Finally, an energy-dependent phase factor, which we could call the `Floquet-Tomasch phase factor' $e^{iq(\pm 2)R}=e^{\pm i\sqrt{(E\pm 2\omega_0)^2-\Delta^2}R/v_F}$ is also accumulated due to the propagation in the middle superconductor. 

\section{Reduction to a Two Level System}\label{TLS}
\begin{figure*}
	\centering
 \includegraphics[width=\textwidth]{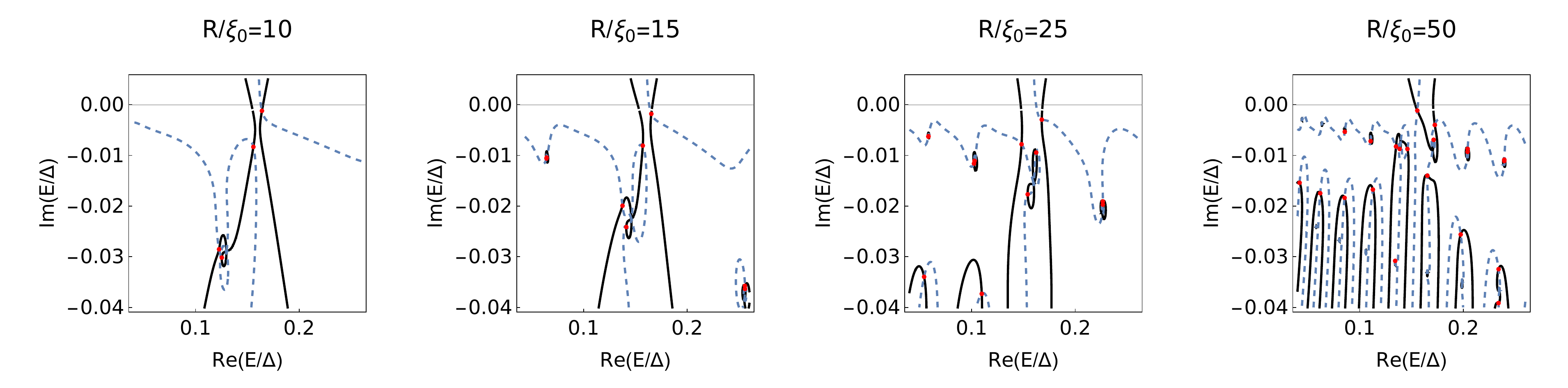}
	\caption{Contour plots of the solutions of Eq. (\ref{transcend}) in the complex plane. The solutions are found at the intersections of $\Re(\det\mathcal{L}_{\mathrm{eff}})=0$ (black lines) and $\Im(\det\mathcal{L}_{\mathrm{eff}})=0$ (blue dashed lines). The roots are marked with red dots. Choice of parameters: $\Gamma_j=0.2\Delta, \omega_0=0.5\Delta.$}
	\label{poles}
\end{figure*} 
With an appropriate transformation, we can show that the basic physics of the system at the regime of interest is that of two resonances coupled through a continuum. The resulting effective Hamiltonian is non-Hermitian, which is a result of the fact that we have focused on the Hamiltonian of a subsystem. The linear operator $\mathcal{L}(0)$ can be transformed into the basis where the matrices $M^0_{1,2}(0)$ of the uncoupled dots are diagonal. We will assume identical dots for simplicity, so that $M^0_{1,2}(0)$ have the same pair of eigenvalues $\pm E_0.$ Details are provided in the Appendix \ref{diagonalization}. We will take into account that due to the particle-hole symmetry of the spectrum the roots of the characteristic polynomial $\det\bqty{\mathcal{L}(0)}=0$ come in pairs (if $E$ is an eigenvalue, so is $-E^{\ast}$). We can then focus on the positive sector of energies only, assuming that the coupling between positive and negative energy states is small. We then find an effective Floquet operator,\\
\begin{widetext}
  \begin{equation}
\begin{split}
     \mathcal{L}_{\mathrm{eff}} &=  \mqty(E-E_0 & 0 \\ 0 & E-E_0)-\cos^2\theta \mqty(\Sigma(0)^{-,11} & \Sigma(0)^{-,14} \\ \Sigma(0)^{-,41} & \Sigma(0)^{-,44}) - \sin^2\theta \mqty(\Sigma(0)^{+,22} & \Sigma(0)^{+,23} \\ \Sigma(0)^{+,32} & \Sigma(0)^{+,33}).
\end{split}
\end{equation}  
\end{widetext}
We see that the parameter $\theta$ (defined in Eq. \ref{angle}) controls the relative strength of the self-energy processes $\Sigma^{\pm},$ where $\Sigma^{-}$ connects the dots through $S_c$ at energies below the gap $E-2\omega_0<-\Delta,$ while $\Sigma^{+}$ connects the dots through quasiparticle propagation in the middle superconductor at energies above the gap $E+2\omega_0>\Delta.$ The parameter $\theta$ itself can be controlled by the voltage and the couplings which change the relative weights of the electronlike and holelike components of the eigenvectors.\par
The resulting effective operator is of the form 
\begin{equation}\label{Leff}
         \mathcal{L}_{\mathrm{eff}} =\mqty(E-E_0+i\gamma & i\gamma_{12} \\ i\gamma_{21} & E-E_0+i\gamma)
\end{equation}
but the $\gamma,$ defined in Eq. (\ref{gamma}), are themselves functions of the energy $E$, the voltage bias $\omega_0$ and the distance $R$ between the resonances. The above relation describes the coupling of two discrete levels initially at $E_0,$ which are coupled through a continuum of states. The overall action of the continuum is, as expected, to add a small shift to $E_0$ equal to the real part of the diagonal self-energy elements and a width equal to their imaginary part. Moreover, the two resonances are then coupled through the non-diagonal elements of the self-energies.\par
The non-diagonal elements that couple the two resonances can be written in a form that makes apparent the dependence on the quartet phase $\phi_q=\phi_a+\phi_b-2\phi_c$
\begin{subequations}
    \begin{align}
    \gamma_{12}&= \alpha e^{i\phi_q}e^{iq(-2)R}-\beta e^{-i\phi_q}e^{iq(2)R} \\
        \gamma_{21}&= \alpha e^{-i\phi_q}e^{iq(-2)R}-\beta e^{i\phi_q}e^{iq(2)R}.    
    \end{align}
\end{subequations}
where the coefficients $\alpha,\beta$ are defined in Eq. (\ref{alphabeta}).\par
\begin{figure*}
    \centering
    \begin{subfigure}{0.4\linewidth}
       \includegraphics[width=\linewidth]{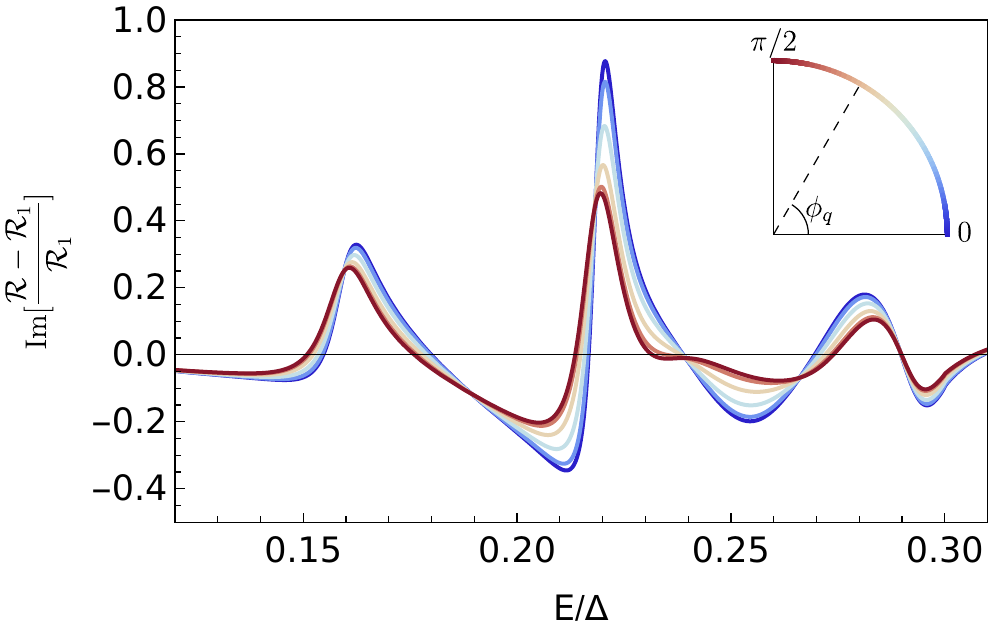} 
           \caption{}
    \label{quartet var}
    \end{subfigure}
     \hspace{1.2cm}
    \begin{subfigure}{0.4\linewidth}
           \includegraphics[width=\linewidth]{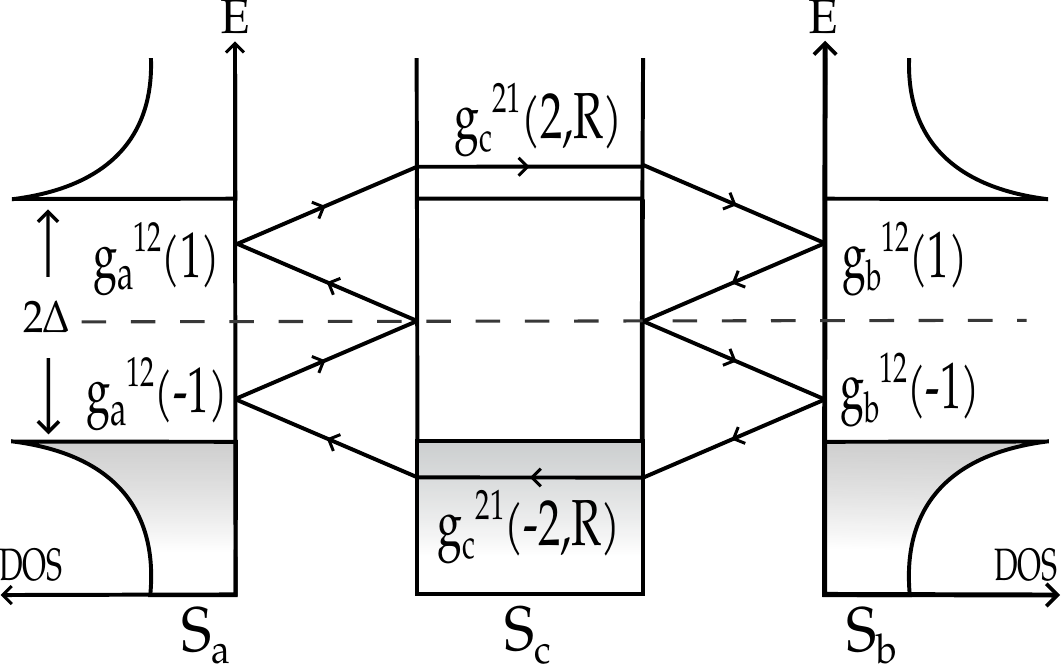} 
               \caption{}
    \label{octet sketch}
    \end{subfigure}
    \caption{(a) Variation of the spectral function $\Im\bqty{\frac{\mathcal{R}-\mathcal{R}_1}{\mathcal{R}_1}}^{11}.$ Different colors represent a different quartet phase,  $\phi_q\in[0,\frac{\pi}{2}],$ as explained in the inset. Other parameters are set to $\omega_0=0.7\Delta$, $\Gamma_j=0.5\Delta$, $R=25\xi_0.$ (b) Sketch of an octet process $\Sigma(0)^{-,14}\times\Sigma(0)^{+,32},$ which couples a hole at energy $E_0$ on dot $1$ to an electron at the same energy on dot $2$ and accumulates a phase $e^{2i\phi_q}.$}
    \label{quartet}
\end{figure*}

 
Finding the resulting eigenvalues due to the coupling between the two resonances requires finding the zeroes of the characteristic polynomial of $\mathcal{L}_{\mathrm{eff}}.$ The characteristic polynomial will be a transcendental equation, generally requiring a numerical solution:
\begin{widetext}
\begin{equation}\label{transcend}
  (E-E_0+i\gamma)(E-E_0+i\gamma)+\alpha^2 e^{2iq(-2)R} +\beta^2 e^{2iq(2)R}-2\alpha\beta\cos2\phi_q e^{iq(2)R}e^{iq(-2)R}=0.
\end{equation}
\end{widetext}
The solutions of the above equation are found numerically and plotted on the complex plane in Fig. \ref{poles}. At small distances, we find two solutions around the initial level $E_0,$ slightly shifted in the complex plane. As the distance between the dots grows, however, there are more solutions which appear around the two initial ones. The number of the solutions increases with the distance since the factors $e^{iq(\pm 2)R}$ become more rapidly oscillating. Figure \ref{poles} shows that the effective model roughly captures the expected behavior, i.e., the number of poles increases with increasing interdot distance, in agreement with Fig. \ref{dos_bijunction} and Fig. \ref{evolution} that were produced by numerically calculating the full operator $\mathcal{L}(0)$.
\paragraph*{Oscillations of the spectral function.}
From Eq. (\ref{Leff}) we can calculate the corresponding effective resolvent operator $\mathcal{R}_{\mathrm{eff}}=\mathcal{L}_{\mathrm{eff}}^{-1}:$
\begin{equation}
\begin{split}
  \mathcal{R}^{11}_{\mathrm{eff}} &=\frac{E-E_0+i\gamma}{(E-E_0+i\gamma)^2+\gamma_{12}\gamma_{21}} \\
  &=\sum_{n=0}^{\infty}\frac{(-\gamma_{12}\gamma_{21})^n}{(E-E_0+i\gamma)^{2n+1}}.   
\end{split}
\end{equation}
For illustrative purposes, we can consider a voltage value $\Delta-E_0<2\omega_0<\Delta+E_0,$ where only the forward self-energy $\Sigma^{+}$ contributes around $E_0>0.$ Then the expression for the resolvent at real energies close to $E_0$ simplifies to:
\begin{equation}
\begin{split}
    \mathcal{R}^{11}_{\mathrm{eff}} &=\sum_{n=0}^{\infty}\frac{(-\beta^2)^n e^{2inq(2)R}}{(E-E_0+i\gamma)^{2n+1}}\\
    &\approx \frac{1}{E-E_0+i\gamma}-\frac{\beta^2 e^{2iq(2)R}}{(E-E_0+i\gamma)^3}+\cdots
\end{split}
\end{equation}
The effect on the spectral function $-\frac{2}{\pi}\Im \mathcal{R}_{\mathrm{eff}}^{11}(E)$ will then be a Breit-Wigner-like resonance around $E_0,$ coming from the first term $(E-E_0+i\gamma)^{-1},$ and smaller oscillations superimposed on the resonance due to the second term on the right-hand side. The spectral function will therefore oscillate as a periodic function of a Floquet-Tomasch factor: $2q(2)R=\frac{2R}{v_F}\sqrt{(E+2\omega_0)-\Delta^2}.$ Since the interdot coupling term $\beta$ is proportional to $\Gamma_a\Gamma_b\Gamma_c,$ we expect that the Floquet-Tomasch oscillations are larger in amplitude when increasing the couplings. At the same time, the width of the resonances given by $\gamma$ is also proportional to the tunnel couplings. Then one expects that the resonances are smeared out with increasing $\Gamma.$ The behavior of the resonances under different couplings and voltages is shown in the Supplemental Material.
\paragraph*{Quartet phase.} In Eq. (\ref{transcend}), the quartet phase $\phi_q$ appears in the last term as $\cos2\phi_q.$ This can be related to ``octet" processes, as discussed in \cite{melin2021}. A sketch of an octet process is shown in Fig. \ref{octet sketch}. Here, Eq. (\ref{transcend}) gives us bounds for the appearance of the octets. At large distances, the Floquet-Tomasch phase factors $e^{iq(\pm 2)R},$ and therefore the corresponding self-energy processes $\Sigma^{\pm}(0),$ are non-zero at the current order in the tunnel couplings only if the condition $\abs{E\pm 2\omega_0}>\Delta$ is satisfied. Then, around a resonance $E\sim E_0>0$ the $e^{iq(2)R}$ term contributes when the voltage is $2\omega_0>\Delta-E_0,$ while the $e^{iq(-2)R}$ term contributes when $2\omega_0>\Delta+E_0.$ As a result, the octet term can only contribute when both processes are present, that is, when $2\omega_0>\Delta+E_0.$ There will therefore be a regime of voltages $\Delta-E_0<2\omega_0<\Delta+E_0,$ where only the forward self-energy $\Sigma^{+}$ contributes around $E_0>0,$ while, making an analogous argument, only the backward self-energy $\Sigma^{-}$ will contribute around $-E_0<0.$ When the voltage is increased above $2\omega_0>\Delta+E_0$ both processes contribute, but with different weights, since the process with the larger absolute value of energy $\abs{E_0\pm 2\omega_0}$ will start to exponentially decay at energies much larger than the gap.\\
In the regime that the octet term is relevant, the quartet phase can nonlocally control the interdot coupling, since $\phi_q$ can be tuned by changing the phase $\phi_b$ across the second junction, while measuring the spectrum on the first.  Equation (\ref{transcend}) suggests that the amplitude of oscillations is enhanced at $\phi_q=0,$ and minimized at $\phi_q=\pi/2.$ Moreover, the quartet phase does not significantly affect the frequency of oscillations, which is rather a function of the energy, the voltage, and the interdot distance. These observations are verified numerically in Fig. \ref{quartet var} that shows the variation of the spectral function of the bijunction with respect to the spectral function of the single junction, $\Im[\frac{\mathcal{R}-\mathcal{R}_1}{\mathcal{R}_1}]^{11},$ calculated numerically for different quartet phases. The numerical calculation is performed without making any approximations, i.e., by calculating the operator $\mathcal{L}$ using Eq. (\ref{chain}). It is worth noting that Fig. \ref{quartet var} implies that the Floquet-Tomasch oscillations will not be smeared out if an average is taken over the quartet phase. The oscillations should therefore be observable even if the quartet phase should drift with time.

\section{Conclusions}\label{conclusions}
\vspace{-10pt}
We have studied two driven superconducting quantum dots connected to a common superconductor. In the limit where the superconductor is long and subgap transport is governed by MARs, we showed that a long-range coupling develops between the dots. By mapping the initial time-periodic problem to a static tight-binding model where time is traded for an extra Floquet dimension, we obtained expressions in continued fraction form for the resolvent operator and the corresponding self-energy. The iterative form of the expressions allows for a fast calculation of the resolvent and can be adapted to other multiterminal configurations. We showed that the system can be described by an effective non-Hermitian model of two resonances coupled through higher-order processes that involve local MARs on each dot, followed by a nonlocal Andreev reflection through the common superconductor at energies above the gap. The induced interdot coupling modifies the Floquet spectrum, producing oscillations in the spectral function. The amplitude of these oscillations can be controlled nonlocally by changing parameters like the phase drop across one of the dots. This amounts to tuning the oscillations with the quartet phase, and we have found bounds for which the quartet phase is involved.\par
It remains to be seen if control of the quartet phase is feasible experimentally at finite voltage bias. This is the topic of a recent preprint, which proposes an interferometric setup sensitive to quartet processes \cite{melininterferometer}. It is therefore an open question whether the quartet phase can be used to control the amplitude of the Floquet-Tomasch oscillations. Regardless, we expect that the oscillations as a function of the energy will not be smeared out, even if we consider a quartet phase that drifts with time. \par
Our approach is relevant for well-defined Floquet resonances on the dots, that is, for tunnel couplings to the reservoirs that are not very large $\Gamma<\Delta,$ and for subgap voltage values. We assumed a large subgap voltage bias $\Delta/2<\omega_0<\Delta$ that allows to simplify the analytical part of this work since at strong driving only a few Floquet harmonics need to be taken into account. However, the mechanism that results in a long-range interdot coupling should exist at smaller subgap voltages as well, although at higher order in the tunnel couplings. Coulomb repulsion $U$ on the dots was assumed small, $U\sim\Gamma<\Delta$, so that the quantum dots can be modeled by single effective levels \cite{yeyati-dot} that we placed in the middle of the superconducting gap. A possible way to include interactions could be to use a master-equation approach \cite{Kosov_2013,PhysRevB.87.155439}, which has the advantage of treating the interactions exactly, but assumes weak coupling to the reservoirs $\Gamma\ll\Delta,$ and therefore does not capture the physics due to MAR processes. An open-quantum system framework that includes both the effect of finite $U$ and of MARs has been proposed \cite{dissipation}, and shows the possibility of engineering the subgap transport through dissipation. It would be interesting to see if the long-range Floquet-Tomasch effect could be similarly engineered.

\appendix
\section{Change of basis}\label{diagonalization}
This section provides details on how to perform a rotation in the basis that diagonalizes $M^0_{1,2}(0).$ For large voltage bias and small tunnel couplings we can make an approximation of $M^0_{1,2}(0)$ by assuming $\abs{E}\ll\omega_0,\Delta:$
\begin{equation}
    M^0_{1,2}\approx \mqty(E\mp\frac{\Gamma_{a,b}\omega_0}{\sqrt{\Delta^2-\omega_0^2}} & -\Gamma_{c_{1,2}} \\ -\Gamma_{c_{1,2}} & E\pm\frac{\Gamma_{a,b}\omega_0}{\sqrt{\Delta^2-\omega_0^2}}).
\end{equation}
The solutions of $\det(M^0_{1,2})=0$ are then 
\begin{equation}
E^{\pm}_{1,2}= \pm \sqrt{\Gamma_{c_1,c_2}^2+\pqty{\frac{\Gamma_{a,b}\omega_0}{\sqrt{\Delta^2-\omega_0^2}}}^2}.
\end{equation}
For simplicity, we will assume that the dots are identical, meaning that we take the couplings $\Gamma_a=\Gamma_b$ and $\Gamma_{c_{1,2}}=\Gamma_c.$ Then, the two matrices $M^0_{1,2}$ have the same pair of eigenvalues $\pm E_0\equiv\pm \sqrt{\Gamma_{c}^2+\pqty{\frac{\Gamma_{a}\omega_0}{\sqrt{\Delta^2-\omega_0^2}}}^2},$ but different eigenvectors. Indeed, the structure of the matrices $M^0_{1,2}$ means that the corresponding eigenvectors satisfying $M^0_{j}(0)\psi_j^{\pm}=0$ can be parametrized as\\
\begin{subequations}
\begin{align}
 	&\psi_1^+=\mqty(\cos\theta \\ \sin\theta),\qquad \psi_1^-=\mqty(-\sin\theta \\\cos\theta),\\
   	&\psi_2^+=\mqty(\sin\theta \\\cos\theta),\qquad \psi_2^-=\mqty(-\cos\theta \\ \sin\theta).
\end{align}
\end{subequations}
We see that the `electron' and `hole' components of the eigenvectors are, in fact, reversed. Moreover, we can derive a simple relation for the angle $\theta$ involving the tunnel couplings and the voltage frequency $\omega_0:$
\begin{equation}\label{angle}
    \theta=\frac{1}{2}\arctan(\frac{\Gamma_c}{\Gamma_a}\frac{\sqrt{\Delta^2-\omega_0^2}}{\omega_0}).
\end{equation}
Within our approximation that $\frac{\Delta}{2}<\omega_0<\Delta,$ we can deduce from the above relation that the principal value of the angle is $\theta\in [0,\pi/4)$. This angle therefore controls the electron/hole content of the eigenvectors.\par
We define change of basis matrices $P(\theta)$ and $Q(\theta)$ that diagonalize $M^0_{1,2}:$
\begin{equation}
	\begin{split}
	&M^0_1(0) =P(\theta)DP(\theta)^{-1}\\
    &=\mqty(\cos\theta & -\sin\theta \\ \sin\theta &\cos\theta)\mqty(E-E_0 & 0 \\0 &E+E_0)\mqty(\cos\theta & \sin\theta \\ -\sin\theta &\cos\theta), \\
	&M^0_2(0)=Q(\theta)DQ(\theta)^{-1}\\
    &=\mqty(\sin\theta & -\cos\theta \\ \cos\theta &\sin\theta)\mqty(E-E_0 & 0 \\0 &E+E_0)\mqty(\sin\theta & \cos\theta \\ -\cos\theta &\sin\theta).
	\end{split}
\end{equation}
Using these, we can transform the initial operator $\mathcal{L}(0)$ on the basis of $\psi_1^+,\psi_1^-, \psi_2^+,\psi_2^-:$
\begin{widetext}
\begin{equation}
	\begin{split}
	\widetilde{\mathcal{L}} &=\mqty(P(\theta)^{-1} & 0 \\ 0 & Q(\theta)^{-1})\mathcal{L}(0)\mqty(P(\theta) & 0 \\ 0 & Q(\theta)) \\
	&=\mqty(P(\theta)^{-1}M^0_1(0)P(\theta) & 0\\ 0 &Q(\theta)^{-1}M^0_2(0)Q(\theta)) -\mqty(P(\theta)^{-1} & 0 \\ 0 & Q(\theta)^{-1})\Sigma(0) \mqty(P(\theta) & 0 \\ 0 & Q(\theta)).
	\end{split}
\end{equation}
By permutation of the basis vectors $\psi_1^-\rightleftharpoons \psi_2^+,$ we can rewrite $\widetilde{\mathcal{L}}$ in order to make apparent the two blocks which correspond to positive and negative eigenvalues. To lowest order of perturbation in the tunnel couplings, we can neglect the non-diagonal blocks in $\widetilde{\mathcal{L}}.$ This amounts to neglecting the coupling between positive and negative eigenvalue sectors. For the upper-left block of $\widetilde{\mathcal{L}}$ we then obtain
\begin{equation}
\begin{split}
     \widetilde{\mathcal{L}}^{++} &=  \mqty(E-E_0 & 0 \\ 0 & E-E_0)-\cos^2\theta \mqty(\Sigma(0)^{-,11} & \Sigma(0)^{-,14} \\ \Sigma(0)^{-,41} & \Sigma(0)^{-,44}) - \sin^2\theta \mqty(\Sigma(0)^{+,22} & \Sigma(0)^{+,23} \\ \Sigma(0)^{+,32} & \Sigma(0)^{+,33}).
\end{split}
\end{equation}
\end{widetext}
The resulting effective operator is of the form 
\begin{equation}
         \mathcal{L}_{\mathrm{eff}} =\mqty(E-E_0+i\gamma_1 & i\gamma_{12} \\ i\gamma_{21} & E-E_0+i\gamma_2).
\end{equation}
Explicitly, the diagonal components of the self-energy matrices will add a finite lifetime to the discrete levels at $E_0,$ given by:
\begin{subequations}\label{gamma}
    \begin{align}
    -i\gamma_1 &= \cos^2\theta \cdot g^{12}_a(-1)\bqty{\frac{1}{M^0(-2)}}^{22}g^{21}_a(-1) \nonumber\\ 
    &+ \sin^2\theta \cdot g^{21}_a(1)\bqty{\frac{1}{M^0(2)}}^{11}g^{12}_a(1), \\
    -i\gamma_2 &= \cos^2\theta \cdot g^{21}_b(-1)\bqty{\frac{1}{M^0(-2)}}^{33}g^{12}_b(-1) \nonumber\\ 
    &+ \sin^2\theta \cdot g^{12}_b(1)\bqty{\frac{1}{M^0(2)}}^{44}g^{21}_b(1),
  \end{align}
\end{subequations}  
and $\gamma_1=\gamma_2\equiv\gamma$ for identical dots. The non-diagonal components will couple the two resonances
\begin{subequations}
    \begin{align}
          \gamma_{12}&= \alpha e^{i\phi_q}e^{iq(-2)R}-\beta e^{-i\phi_q}e^{iq(2)R}, \\
        \gamma_{21}&= \alpha e^{-i\phi_q}e^{iq(-2)R}-\beta e^{i\phi_q}e^{iq(2)R},  
    \end{align}
\end{subequations}
where
\begin{subequations}\label{alphabeta}
    \begin{align}
         \alpha =&\frac{\Gamma_a\Gamma_b\Gamma_c\Delta^3\cos(k_FR)\cos^2\theta}{\bqty{\Delta^2-(\omega_0-E)^2}\sqrt{(2\omega_0-E)^2-\Delta^2}} \nonumber\\
         &\times\bqty{\frac{1}{M^0_1(-2)}}^{22}\bqty{\frac{1}{M^0_2(-2)}}^{11},\\
        \beta =&\frac{\Gamma_a\Gamma_b\Gamma_c\Delta^3\cos(k_FR)\sin^2\theta}{\bqty{\Delta^2-(\omega_0+E)^2}\sqrt{(2\omega_0+E)^2-\Delta^2}} \nonumber\\
        &\times\bqty{\frac{1}{M^0_1(2)}}^{11}\bqty{\frac{1}{M^0_2(2)}}^{22}.
    \end{align}
\end{subequations}

\bibliography{bibliography}

\end{document}


\title{Long-range coupling between superconducting dots induced by periodic driving: Supplemental Material}

\author{Andriani Keliri}\email{akeliri@lpthe.jussieu.fr}
 \affiliation{Laboratoire de Physique Th\'{e}orique et Hautes Energies,
Sorbonne Universit\'{e} and CNRS UMR 7589, 4 place Jussieu, 75252 Paris Cedex 05, France}
\author{Beno\^{i}t Dou\c{c}ot}
 \affiliation{Laboratoire de Physique Th\'{e}orique et Hautes Energies,
Sorbonne Universit\'{e} and CNRS UMR 7589, 4 place Jussieu, 75252 Paris Cedex 05, France}

\maketitle

\section{Single junction}
For a single dot coupled to two superconductors with $V_a=-V, V_c=0,$ we denote the Floquet chain operator by $\mathcal{L}_1(m)=M^0_1(m)-\Sigma^{+}_1(m)-\Sigma^{-}_1(m),$ and its inverse by the resolvent operator $\mathcal{R}_1=\mathcal{L}_1^{-1},$ where 
\begin{equation}
M^0_1(m)=E+m\omega_0-g_c(m)-\mqty(g_a^{11}(m-1) & 0 \\ 0 & g_a^{22}(m+1)),
\end{equation}
and the self-energy matrices are:
\begin{subequations}
\begin{align}
    \Sigma_1^{+}(m) &=\mqty(0 & 0 \\ 0 & g^{21}_a(m+1)\bqty{\frac{1}{M^0_1(m+2)-\Sigma^{+}_1(m+2)}}^{11}g^{12}_a(m+1)),    \\
    \Sigma_1^{-}(m) &=\mqty(g^{12}_a(m-1)\bqty{\frac{1}{M^0_1(m-2)-\Sigma^{-}_1(m-2)}}^{22}g^{21}_a(m-1) & 0 \\ 0 & 0).    
\end{align}
\end{subequations}
The self-energy will therefore add dissipation to the diagonal elements of $M^0_1$ due to MAR processes. To simplify things, if we consider a voltage such that $\frac{\Delta}{2}<\omega_0<\Delta,$ we see that if an initial ABS level is close to an energy $\abs{E}\ll \Delta,$ then two Andreev reflections would be enough to connect the level to the continuum of states above or below the gap, by absorption or emission of virtual photons of energy $2\omega_0.$ In this case the self-energies will have an imaginary part proportional to $g_a^{12}(\pm 1)g_c^{22,11}(\pm 2)g_a^{21}(\pm 1)$ and therefore are of order $\order{\Gamma_a^2 \Gamma_c}$ in the tunnel couplings. This imaginary part will push the zeroes of $M^0_1$ below the real axis. 
\begin{figure}
    \centering
    \includegraphics[width=0.42\linewidth]{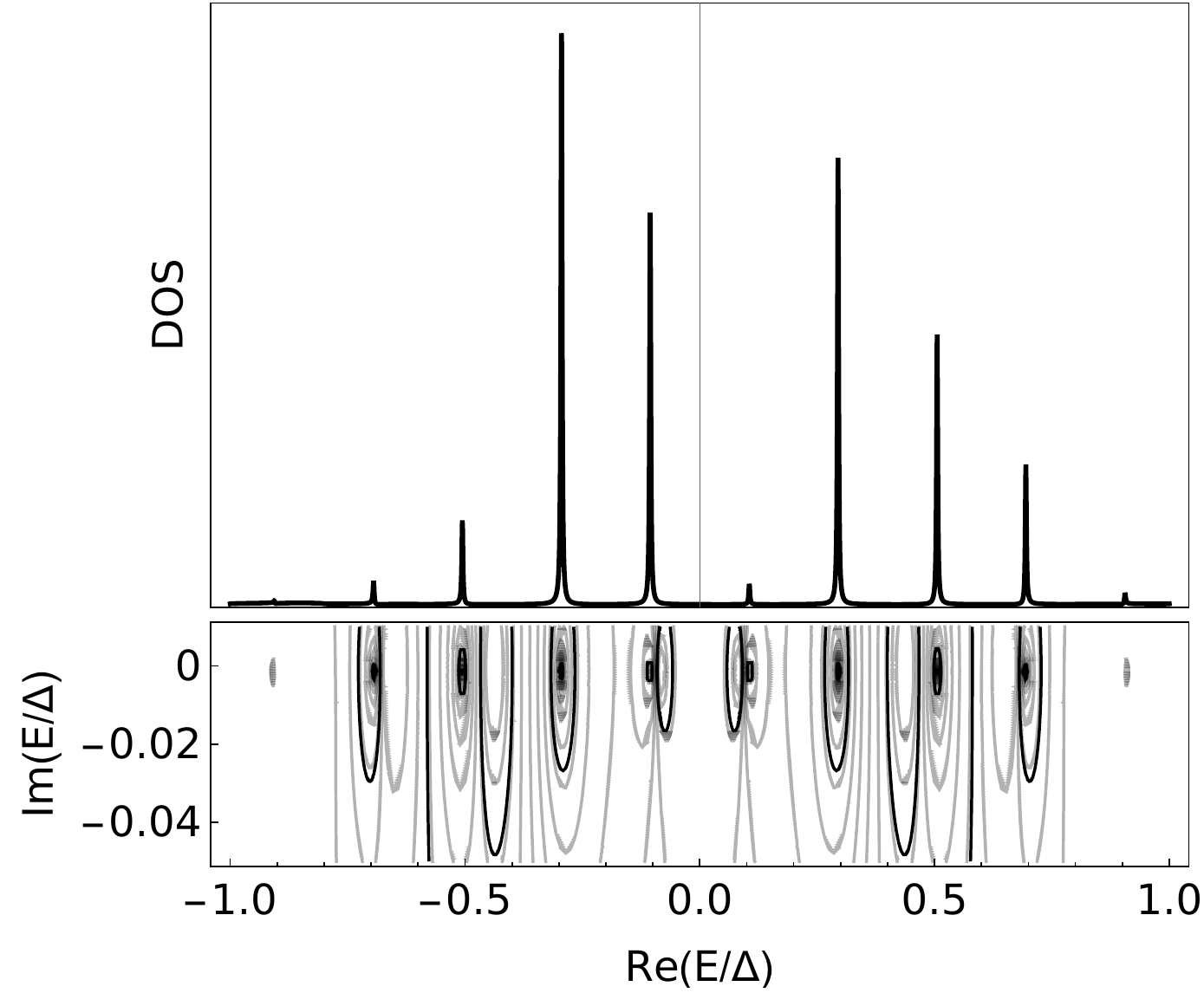}\hspace{20pt}
    \includegraphics[width=0.42\linewidth]{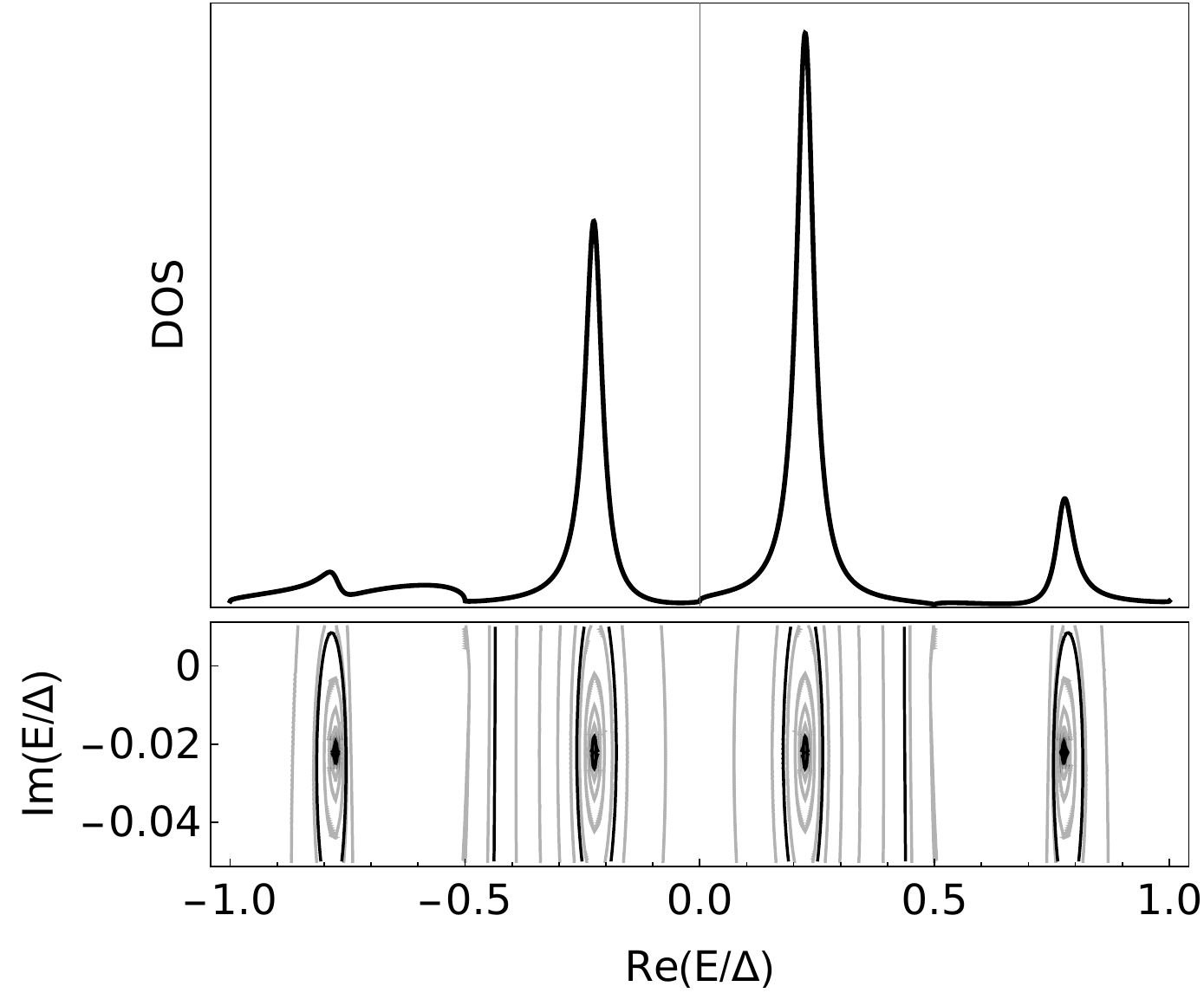}
    \caption{Multiple Andreev reflections turn the initial ABS on the dot into resonances with a finite width. Due to the time-periodicity Floquet replicas appear. Contour plot of $\log\abs{\det\mathcal{L}_1(0)}$ showing the zeroes of $\det\mathcal{L}_1$ in the complex plane (lower panel) and corresponding spectral function $-\Im\mathcal{R}^{11}_1$ (peaks in the upper panel). (a) Drive frequency $\omega_0=\Delta/5,$ and (b) $\omega_0=\Delta/2.$ The couplings are set to $\Gamma_{a,c}=\Delta/2$ on both plots. The height of the peaks should not be compared across different plots.}
    \label{singlejunction}
\end{figure}
Figure \ref{singlejunction} shows the zeroes of $\det\mathcal{L}_1(0)$ and the corresponding peaks in the spectral function given by the imaginary part of the resolvent operator $-\frac{2}{\pi}\Im\mathcal{R}^{11}_1(0)$; if a zero of $\det\mathcal{L}_1(0)$ appears at an energy $E_0-i\gamma,$ then the spectral function will have a peak around the energy $E_0,$ with a width given by the imaginary part $\gamma.$ 

\section{Variation of the spectral function due to Floquet-Tomasch effect}
\begin{figure}[h]
    \centering
    \begin{subfigure}{\linewidth}
    \includegraphics[width=\linewidth]{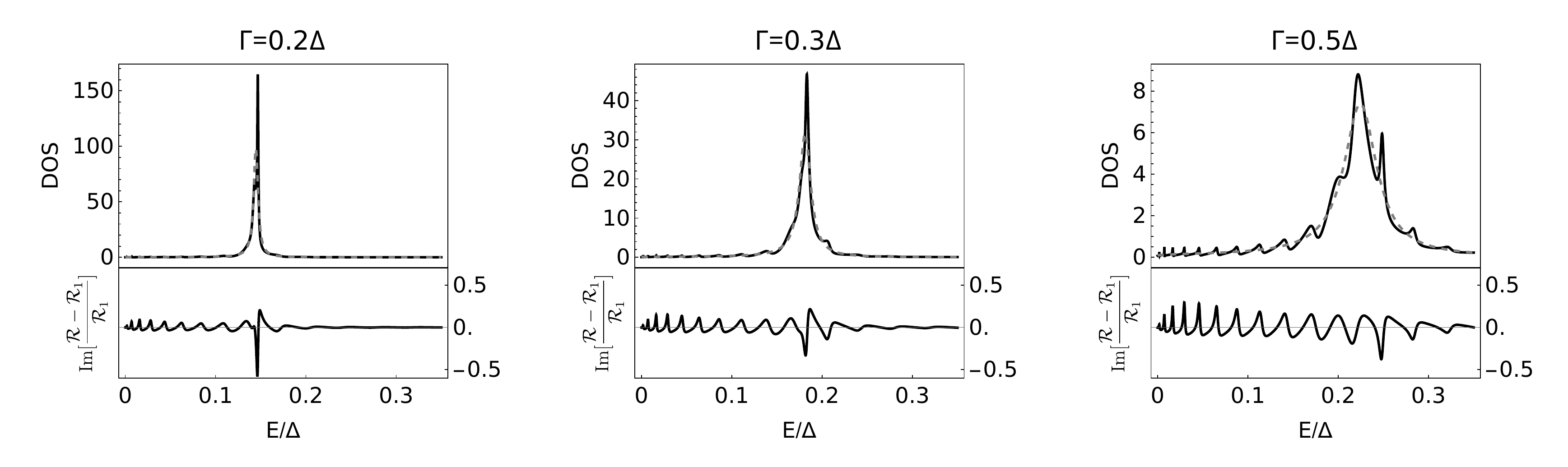} 
    \caption{}
        \vspace{5pt}
    \end{subfigure}
   \begin{subfigure}{\linewidth}
     \includegraphics[width=\linewidth]{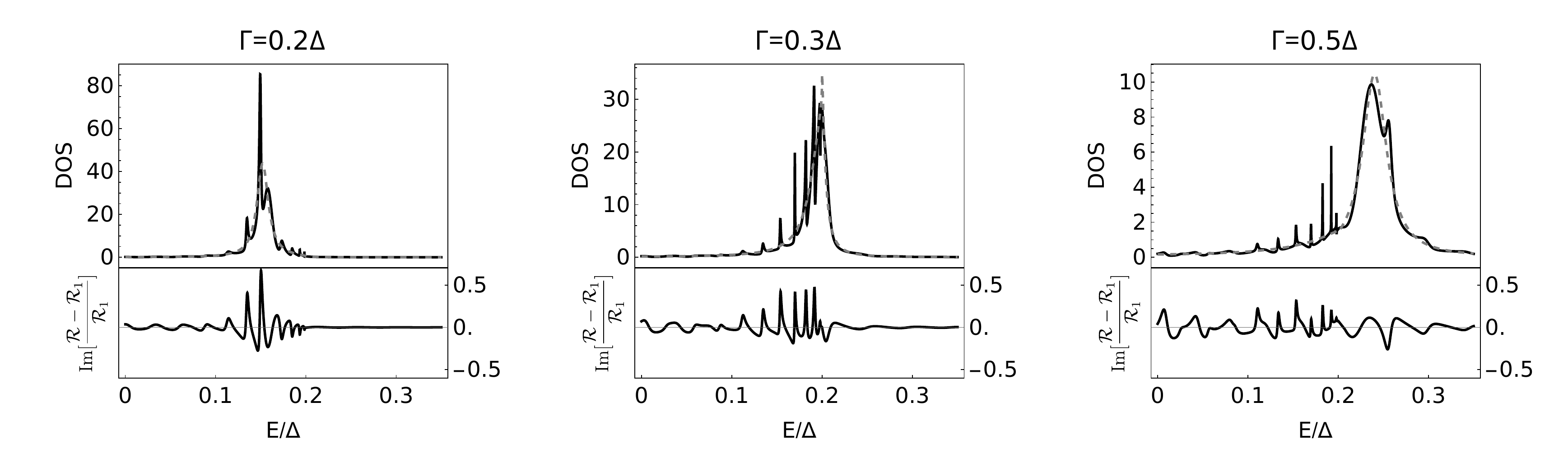} 
    \caption{}
    \vspace{5pt}
   \end{subfigure}
    \begin{subfigure}{\linewidth}
     \includegraphics[width=\linewidth]{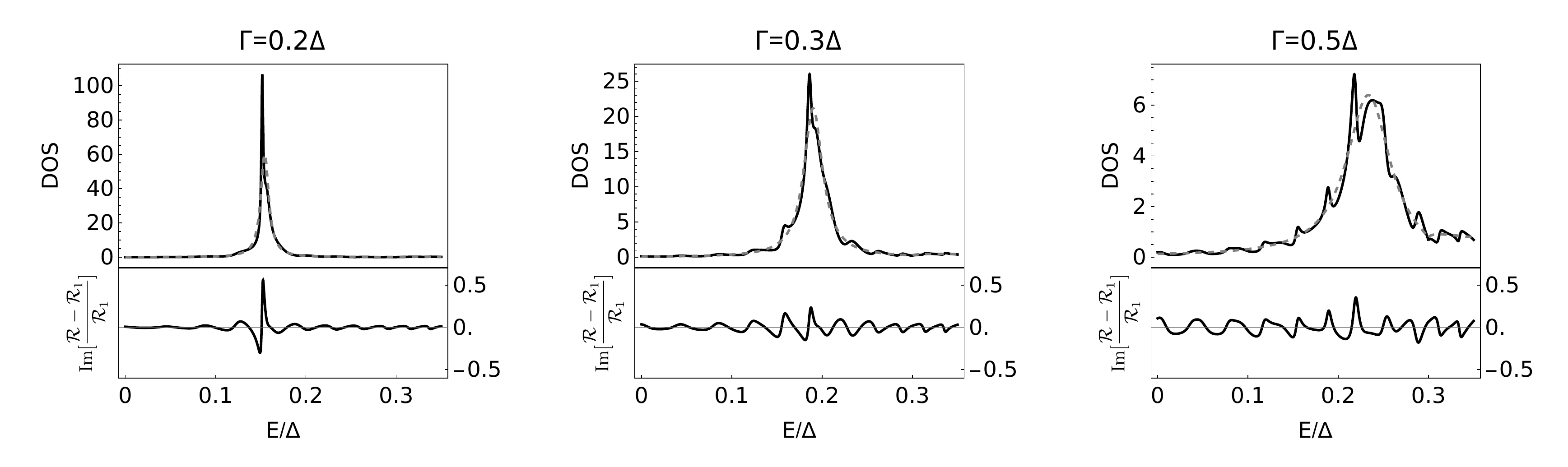} 
    \caption{}
   \end{subfigure}
    \caption{DOS of the single junction $-\frac{2}{\pi}\Im\mathcal{R}_1^{11}$ (grey line, dashed) compared to the bijunction $-\frac{2}{\pi}\Im\mathcal{R}^{11}$ (black line) and variation of the spectral function $\Im\bqty{\frac{\mathcal{R}-\mathcal{R}_1}{\mathcal{R}_1}}^{11}$ (lower panels). The distance between the dots is set to $R=50\xi_0$. The drive frequency is (a) $\omega_0=0.5\Delta,$ (b) $\omega_0=0.6\Delta,$ and (c) $\omega_0=0.7\Delta.$}
    \label{oscillations}
\end{figure}

In the regime of large separation between the dots of the bijunction $R\gg\xi_0,$ the spectral function of dot $1$ will be modified due to the coupling to dot $2.$ The resolvent of the bijunction can be written so as to separate the terms corresponding to the uncoupled single junctions $\mathcal{L}_{1,2}$ and the modification due to the coupling between them $\delta\mathcal{L}$
\begin{equation}
\mathcal{R}(0)=\bqty{\mqty*(\mathcal{L}_1(0) & 0 \\ 0 & \mathcal{L}_2(0))-\mqty*(\delta\mathcal{L}_1 & \delta\mathcal{L}_{12} \\ \delta\mathcal{L}_{21} & \delta\mathcal{L}_2)}^{-1}.
\end{equation}
The spectral function on dot $1$ will depend on the $\mathcal{R}^{11}$ element of the resolvent, which can be approximated by
\begin{equation}\label{correction}
\begin{split}
    \bqty{\mathcal{R}}^{11} &=\bqty{\frac{1}{\mathcal{L}_1-\delta\mathcal{L}_1-\delta\mathcal{L}_{12}\mathcal{R}_2\delta\mathcal{L}_{21}}}^{-1} \\
    &\approx \bqty{\mathcal{R}_1}^{11} + \bqty{\mathcal{R}_1\delta\mathcal{L}_1\mathcal{R}_1}^{11} + \bqty{\mathcal{R}_1\delta\mathcal{L}_{12}\mathcal{R}_2\delta\mathcal{L}_{21}\mathcal{R}_1}^{11}+\cdots.
\end{split}
\end{equation}
The spectral function on dot $1,$ will consist of peaks corresponding to the single junction resolvent $\mathcal{R}_1$ and of smaller oscillations superimposed on the peaks due to the correction terms on the right hand side of Eq. (\ref{correction}). The variation of the spectral function of the bijunction with respect to the spectral function of the single junction will therefore reveal how the Floquet-Tomasch effect modifies the spectrum. In Fig. \ref{oscillations}, we plot the spectral function of the bijunction $-\Im\mathcal{R}^{11}$ focusing around a resonance at positive energy (upper panels), and the variation of the spectral function $\Im[\frac{\mathcal{R}-\mathcal{R}_1}{\mathcal{R}_1}]^{11}$ showing the Floquet-Tomasch oscillations (lower panel).
The coupling between dot $1$ and dot $2$ is expressed by the block matrices $\delta\mathcal{L}_{12}$ and $\delta\mathcal{L}_{12}.$ These are $2\times 2$ matrices with self-energy elements in the non-diagonal
\begin{equation}
    \delta\mathcal{L}_{12}=\mqty(0 & \Sigma^{-,14}\\ \Sigma^{+,23} & 0),\qand \delta\mathcal{L}_{21}=\mqty(0 & \Sigma^{+,32}\\ \Sigma^{-,41} & 0),
\end{equation}
where
\begin{subequations}\label{chain}
\begin{align}
    \Sigma^{-,14} &= g^{12}_a(-1)\bqty{\frac{1}{M^0(-2)}}^{23}g^{12}_b(-1) \approx-g^{12}_a(-1)\bqty{\frac{1}{M^0_1(-2)}}^{22}g^{21}_c(-2,R)\bqty{\frac{1}{M^0_2(-2)}}^{11}g^{12}_b(-1)\\
    \Sigma^{+,23} &= g^{21}_a(1)\bqty{\frac{1}{M^0(2)}}^{14}g^{21}_b(1) \approx -g^{21}_a(1)\bqty{\frac{1}{M^0_1(2)}}^{11}g^{12}_c(2,R)\bqty{\frac{1}{M^0_2(2)}}^{22}g^{21}_b(1)\\
    \Sigma^{+,32} &= g^{12}_a(1)\bqty{\frac{1}{M^0(2)}}^{41}g^{12}_b(1) \approx-g^{12}_a(1)\bqty{\frac{1}{M^0_2(2)}}^{22}g^{21}_c(2,R)\bqty{\frac{1}{M^0_1(2)}}^{11}g^{12}_b(1)\\
    \Sigma^{-,41} &= g^{21}_a(-1)\bqty{\frac{1}{M^0(-2)}}^{32}g^{21}_b(-1) \approx -g^{21}_a(-1)\bqty{\frac{1}{M^0_2(-2)}}^{11}g^{12}_c(-2,R)\bqty{\frac{1}{M^0_1(-2)}}^{22}g^{21}_b(-1).
\end{align}
\end{subequations}
The above self-energy elements are proportional to $\Gamma_a\Gamma_b\Gamma_c.$ We therefore expect that the Floquet-Tomasch oscillations are larger in amplitude when increasing the couplings. Indeed, Fig. \ref{oscillations} shows that at small couplings $\Gamma\ll\Delta$ resonances are sharper and oscillations smaller, and the amplitude of oscillations increases with increasing tunnel couplings. However, since the coupling of the resonances to the dissipating continua is also proportional to the tunnel couplings, we expect resonances to become broader with increasing $\Gamma.$
